\documentclass{article}
\usepackage{ssd08}

\usepackage{cite}
\usepackage{graphicx,overpic}
\usepackage{amssymb,amsmath,bm}
\usepackage{hyperref}

\title{A basic framework for the cryptanalysis of digital chaos-based cryptography}

\name{David Arroyo$^1$, Gonzalo Alvarez$^1$ and Veronica
Fernandez$^1$\thanks{The work described in this paper was supported
by \textit{Ministerio de Educaci\'on y Ciencia of Spain}, research
grant SEG2004-02418, \textit{Ministerio de Ciencia y Tecnolog\'{i}a}
of Spain, research grant TSI2007-62657 and \textit{CDTI, Ministerio
de Industria, Turismo y Comercio of Spain} in collaboration with
Telef\'onica I+D, Project SEGUR@ with reference CENIT-2007 2004.}}
\address{$^1$Instituto de F\'{\i}sica Aplicada, Consejo Superior de Investigaciones
Cient\'{\i}ficas,\\
Serrano 144---28006 Madrid, Spain}

\begin{document}

\maketitle

\begin{abstract}
Chaotic cryptography is based on the properties of chaos as source
of entropy. Many different schemes have been proposed to take
advantage of those properties and to design new strategies to
encrypt information. However, the right and efficient use of chaos
in the context of crypto\-graphy requires a thorough knowledge about
the dynamics of the selected chaotic system. Indeed, if the final
encryption system reveals enough information about the underlying
chaotic system it could be possible for a cryptanalyst to get the
key, part of the key or some information somehow equivalent to the
key just analyzing those dynamical properties leaked by the
cryptosystem. This paper shows what those dynamical properties are
and how a cryptanalyst can use them to prove the inadequacy of an
encryption system for the secure exchange of information. This study
is performed through the introduction of a series of mathematical
tools which should be the basic framework of cryptanalysis in the
context of digital chaos-based cryptography.
\end{abstract}


\vspace*{0.5in}
\vspace*{-\baselineskip}

\section{Introduction}
Chaotic cryptography has been an important research area during the
last two decades. The properties of chaotic systems have been used
in very different ways to build new cryptosystems. All of those
proposals can be classified into two big families, which are analog
chaos-based cryptosystems and digital chaos-based cryptosystems. The
first type of chaotic cryptosystems is based on the chaotic
synchronization technique \cite{Pecora1990}, whereas digital chaotic
cryptosystems are designed for digital computers. The scope of this
paper is related to the last type of chaotic cryptosystems. Since
the performance and accuracy of digital chaotic cryptosytems is a
translation of the dynamical properties of the underlying dynamical
system, the first step in either their design or their analysis
comes from the evaluation of those properties. This assessment has
as main goal the verification that the selected dynamical system
evolves in a chaotic way in the context defined by the encryption
strategy. Furthermore, it must be confirmed the impossibility of
guessing the key or part of the key by the examination of that
chaotic behavior. Indeed, chaotic systems exhibit an underlying
regularity which could represent a problem in the context of chaotic
cryptography. The present work presents a set of mathematical tools
for the analysis of the dynamical properties of chaotic systems and,
consequently, for the detection of design problems in digital
chaos-based cryptosystems. More specifically, this work studies the
relationship between the set defined by the initial conditions and
the control parameter of chaotic systems and their temporal
evolution. From the point of view of the security concerns of
cryptography, this work is a first step to establish a paradigm of
requirements for the adequate association of chaotic systems and
encryption architectures.

\section{Chaos and cryptography}
The use of chaotic systems in cryptography is motivated by the
similarity between the needs of cryptography and the main
characteristics of chaos. Indeed, every \emph{cryptosystem}, i.e.,
every encryption system performs the transformation of an input (the
\emph{plaintext}) into an output (the \emph{ciphertext} or
\emph{cryptogram}) in such a way that the different inputs and
outputs of the systems are statistically independent. Moreover, this
transformation depends on an external parameter called \emph{key} of
the cryptosystem. If a cryptosystem has been designed in a correct
way, then its output is statically independent from the key and from
the input. In other words, a good cryptoystem shows \emph{diffusion}
and \emph{confusion} properties \cite{shannon49}. The confusion
property means that the temporal evolution of the output of the
cryptosystem is statistically independent from the input and from
the key. Chaotic systems shows an ergodic behavior and,
consequently, possess inherently the confusion property with respect
to the initial conditions and the control parameter(s). On the other
hand, the diffusion property implies that an small change in the
input of the cryptosystem is translated into a large modification in
the output. Chaotic systems are characterized by a high sensitivity
to initial conditions and the control parameter(s) and thus chaos
can also be used as a source of diffusion. Nevertheless, chaotic
systems are sustained by an underlying regularity by means of the
existence of dense unstable periodic orbits which further constitute
the \emph{skeleton} of chaos \cite[p. 413]{hilborn}. Furthermore,
chaotic systems generally considered for cryptographic applications
are defined in a parametric way and not all the values of the
control parameter(s) lead to a chaotic behavior. As a result, the
design of a secure and efficient chaos-based cryptosytem requires
the selection of an encryption architecture that conceals the
underlying regularity of the chosen chaotic system and, at the same
time, guarantees that the control parameter(s) used in the
encryption and decryption processes are those which drive the
dynamical system into a chaotic behavior. Concerning this last
point, it is highly recommendable to use dynamical systems with a
large and continuous set of values for the control parameter(s) such
that evolution in a chaotic way is guaranteed. According to all this
considerations, the cryptanalysis work can be only successful when a
chaotic cryptosystem leaks too much information about the dynamics
of the underlying chaotic system or when the selection of the
control parameter(s) is not done conveniently.

\section{Detection of non-chaotic behavior}
The scope of this paper is digital chaos-based cryptography and thus
the next work is focused on dynamical systems defined in discrete
time. These dynamical system are also called \emph{maps} and are
mathematically defined by a difference equation
\begin{equation}
 x_{k+1}=g(\mu,x_k),
\end{equation}
where $x, g(x) \in U \in \mathbb{R}^n$, being $U$ the phase space,
and $\mu$ is an array of control parameters in $V \subset
\mathbb{R}^p$. This section is concerned with the definition of
mathematical tools to decide whether a digital chaotic cryptosystem
defines a good framework for the selection of the set of control
parameters of the underlying chaotic map.

\subsection{Bifurcation diagram}
\begin{figure}[!htbp]
    \includegraphics{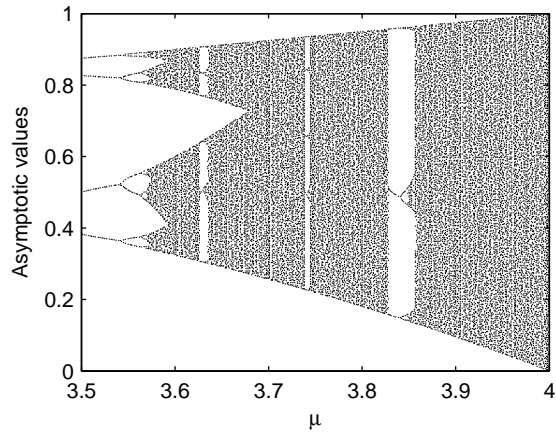}
    \caption{Bifurcation diagram of the logistic map.}
    \label{figure:bifurcationDiagram}
\end{figure}

A first method to detect problems related to the definition of the
method of selection of values for the array of control parameters is
based on the evaluation on the asymptotic temporal evolution. The
temporal evolution or \emph{orbit} of a discrete-time dynamical
system is determined for a certain initial condition and a certain
set of values for the array of control parameters as
\begin{equation}
    \gamma(\mu, x_0)= \left\{x_0,x_1,\ldots,x_{n}\right\}.
\end{equation}
If the considered dynamical system is a chaotic map, then the orbit
derived from any initial condition covers the whole phase space.
This characteristic can be verified by plotting $\gamma(\mu, x_0)$
for $x_0$ selected randomly in $U$ and $\mu$ taking all the values
in $V$. This representation is called \emph{bifurcation diagram} and
can be used to detect values of $\mu$ not leading to chaos. As an
example, in Fig.~\ref{figure:bifurcationDiagram} the bifurcation
diagram for the logistic map is shown identifying those
\emph{inadequate} values of $\mu$. When considering a chaotic
cryptosystem, the possibility of selecting those values of $\mu$
associated to a non-chaotic behavior implies a degradation of the
performance of the cryptosystem and can be used by a cryptanalyst
\cite{alvarez03b,arroyo07c}.

\subsection{The Lyapunov exponent}
\begin{figure}[!htbp]
    \begin{center}
    \begin{overpic}{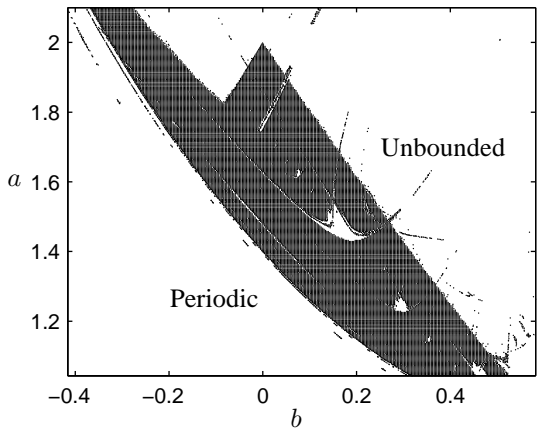}
    \put(50,0){$b$}
    \put(3,40){$a$}
    \put(30,20){Periodic}
    \put(65,45){Unbounded}
    \end{overpic}
    \caption{Determination of the chaotic region of the H\'enon by means of the Lyapunov exponent.}
    \label{figure:chaosHenon}
    \end{center}
\end{figure}

One important rule for the design of chaotic cryptosystems is the
accurate definition of the key space \cite[Rule 5]{Alvarez06a}.
Therefore, cryptanalysis is also concerned with the evaluation of
the preciseness of the definition of the key space. A very useful
tool for this assessment is the Lyapunov exponent or LE. Indeed, the
LE quantifies the local divergence of a dynamical system in such a
way that a positive value of the LE informs of a chaotic behavior.
More rigorously, for a phase space defined in $\mathbb{R}^m$ there
exist $m$ Lyapunov exponents and chaos exists when the largest
Lyapunov exponent is positive. The H\'enon map is a two dimensional
dynamical system and the control parameter vector is given by
$\mu=[a,b]$. If the H\'enon is considered for cryptographic
applications, then the encryption architecture must assure that the
selection of $a$ and $b$ during the encryption/decryption procedure
always guarantees a positive LE (see Fig.~\ref{figure:chaosHenon}).
Otherwise, a degradation in the performance of the cryptosystem is
observed as it occurs in the cryptosytesm analyzed in
\cite{alvarez04c,li05b,arroyo07a}

\section{Exploiting the leaking of the underlying chaotic system regularity}
Most chaos-based cryptosystems used either the initial conditions or
the control parameter(s) or both as key. Therefore, the security of
those proposals is very dependent on the possibility of estimating
the control parameter(s) and/or the initial conditions from the
information provided by the cryptosystem. The success of this
estimation process is something to consider in the design of a new
cryptosystem. Indeed, if it is known how much information can be
provided to a cryptanalyst without exposing our encryption system,
then it is possible to build a secure but also more efficient
cryptosystem. The information that can be used by an attacker or
cryptanalyst is either part of an orbit or the transformation of
part of an orbit of the underlying chaotic system. Actually, that
part of an orbit or the transformation of the part is associated to
a certain initial condition and control parameter(s) value.
Consequently, the cryptanalyst work is focused on looking for
one-to-one relationship between those samples of orbits or those
transformation of the samples of orbits. Next some mathematical
procedures for this estimation task are shown.

\subsection{Study of histograms and the return map}
\begin{figure}[!htbp]
    \center
    \includegraphics{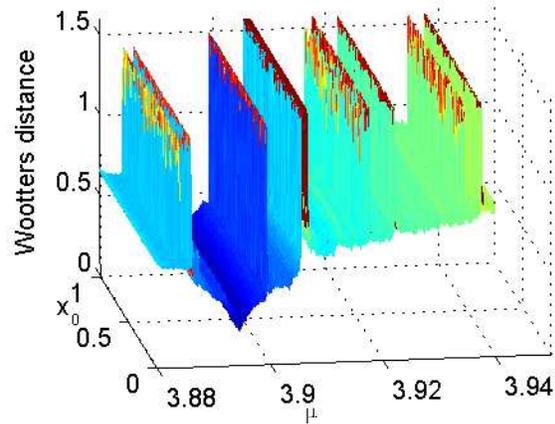}
    \caption{Estimation of the control parameter of the logistic map by means of the Wootters' distance. The value of $\mu$
    and the initial condition $x_0$ used in the generation of the orbit were $\mu=3.8947192$ and $x_0=0.99842379$.}
    \label{figure:wootters}
\end{figure}

In some digital chaos-based cryptosystems the result of the
encryption procedure, i.e., the different cryptograms are given
directly by the orbit of some chaotic system \cite{ling07} or by the
sampling process on the orbit of a chaotic system
\cite{alvarez99,pisarchik06}. In the first case, the cryptosystem
shows a clear security problem, since chaotic maps are defined using
a difference equation which implies that two consecutive values of
an orbit can be used to estimate the control parameter value
\cite{arroyo07c}. In the second situation, the orbit is sampled and
thus it is not possible a direct estimation of the control
parameter(s) as before. However, it is possible to associate those
sampling values with the control parameter(s) by either the study of
the return map or by an statistical treatment. Indeed, according to
Kerckhoffs' principle \cite[p. 14]{menezes:book97}, the chaotic map
used for a certain digital chaos-based cryptosystem is known by
anyone and, consequently, a cryptanalyst knows and can study its
return map. In \cite{skrobek08,arroyo08a}, for example, this study
is used to estimate the control parameter of the logistic map by
means of the maximum value of its return map. On the other hand, the
statistical treatment of sampled orbits is based on the possibility
of reconstructing the probability distribution functions of the
orbits by means of the histograms of their sampled versions. If the
chaotic map used for encryption has good characteristics from the
statistical point of view, then it is not possible to infer the
value of the control parameter(s) from the obtained histograms.
Conversely, a bad chaotic map generates histograms depending on the
value of the control parameter(s). A way to use this dependency for
cryptanalysis is by the Wootters' distance \cite{mlmp:qjsd04}
between different histograms. In this sense, the histogram obtained
from the sampled orbit leaked by the cryptosystem is compared to the
histograms generated from a set large enough of control
parameter(s). The control parameter that generates the closest
histogram to the one obtained from the sampled orbit is considered
as an estimation of the control parameter used for the encryption.
Figure \ref{figure:wootters} illustrates the success of this
estimation method for the logistic map.

\begin{figure}[!htbp]
    \center
    \includegraphics[width=8cm,height=8cm]{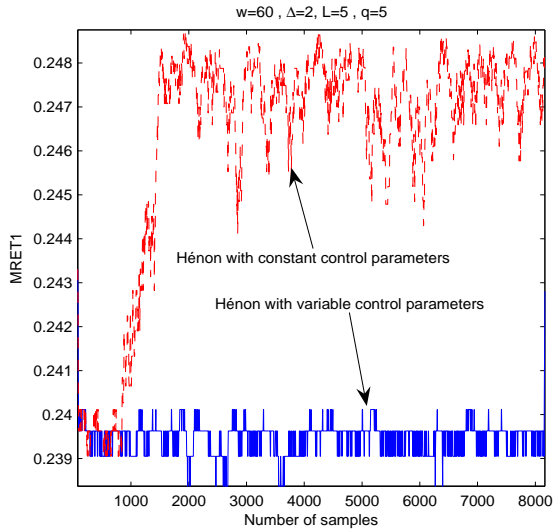}
    \caption{First level of detail of the MultiResolution Entropy of Tsallis of the H\'enon map with fixed and variable
    control
    parameters. See \cite{gamero97} for more details about the procedure.}
    \label{figure:changesHenon}
\end{figure}

\subsection{Analysis of entropy}
In the context of chaos-based cryptography the space of cryptograms
has an entropy determined mostly by the underlying chaotic system.
Therefore, it is very important to verify that the measure of
entropy obtained through the cryptograms does not allow to guess
segments of the plaintext or to get a one-to-one relationship with
respect to the control parameter(s) of the underlying chaotic
system. In this sense, when selecting the chaotic map and the
encryption architecture, it is highly advisable to evaluate first
the entropy of the orbits generated from the map. If this analysis
is not done, then it could be possible for the cryptanalyst to build
an attack upon different measures of entropy depending on the kind
of information contained in the cryptograms. This is the case of the
attack performed in \cite{alvarez03a} on Shannon's measure of
entropy. Other measures of entropy that could be useful for
cryptanalysis are the topological entropy \cite{adler65} and the
wavelet entropy \cite{Rosso01}. Related to the topological entropy,
in the context of digital chaos-based cryptosystems, i.e., when
dealing with chaotic maps it is advisable to work with
approximations of the topological entropy based on the study of
permutations \cite{bandt02a}, which allows to develop a theory of
discrete chaos \cite{kocarev06} with clear interest for
cryptanalysis \cite{amigo07b}. On the other hand, the use of
wavelets for the determination of measures of entropy can be an aid
in the analysis of the security of chaos-based cryptosystems. For
example, the theory of wavelets and the figures of entropy defined
by Shannon and Tsallis can be combined to define a MultiResolution
Entropy analysis that allows to identify changes in the dynamics of
a certain dynamical system \cite{gamero97} (see
Fig.~\ref{figure:changesHenon}). The detection of this changes can
lead to the recovery of information or the guessing of the key or
part of the key of a chaotic cryptosystem, and it can also be
performed through the study of the permutations of the output
sequence of the cryptosystem \cite{cao04}.

\subsection{Analysis of statistical complexity}
\begin{figure}[!htbp]
    \centering
    \includegraphics{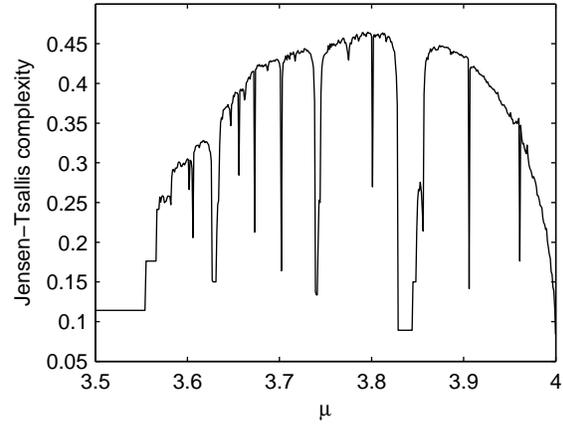}
    \caption{Jensen-Tsallis statistical complexity of the logistic map for different values of $\mu$.}
    \label{figure:complexity}
\end{figure}

Another way of getting benefit from the statistical evaluation of
the output of a chaos-based cryptosystem is based on the notion of
statistical complexity \cite{martin06}. The statistical complexity
measures the level of local divergence that exists in a dynamical
system with respect to the underlying periodic behavior and,
consequently, can be used as a causality indicator. In this sense,
the statistical complexity has been proposed as a test for the
evaluation of the quality of random number generators
\cite{gonzalez05,larrondo05,larrondo06}. Nevertheless, this is not
its only application in the context of cryptanalysis, since it can
also be used to find one-to-one relationships with respect to the
control parameters and, as a result, it could be a hint for control
parameters estimation \cite{rosso07}. As an example, the statistical
complexity of the logistic map with respect to $\mu$ is shown in
Fig.~\ref{figure:complexity} and it proves that the estimation of
$\mu$ is possible through the statistical complexity for certain
intervals inside the definition space of the control parameter.

\subsection{Symbolic dynamics}
The study of the dynamics of unimodal maps through a discretized
version of their phase space is a well known area of study in the
context of the theory of dynamical systems since the seminal
contribution of \cite{metropolis73}. The conclusions derived from
this work has been used for the estimation of the initial condition
and the control parameter values that lead to a certain discretized
version of an orbit associated to an unimodal map
\cite{wu04,physcon07}, which is the base of the cryptanalysis done
in \cite{alvarez03a,arroyo08c,wang05}. However, the theory of
symbolic dynamics, i.e., the study of those discretized versions of
orbits of chaotic maps is not limited to unimodal maps. Indeed, the
possibility of applying this kind of methodology is based on the
underlying periodicity that sustained chaos. In other words, if the
phase space of a chaotic map is partitioned conveniently and a
symbol is assigned to each of the derived intervals, then it is
possible to get the unstable periodic orbits of the system and,
consequently, to reconstruct its dynamics
\cite{kennel03,buhl05,rajagopalan06,buhl07}. This is very
interesting from the point of view of cryptanalysis, since it allows
to estimate the control parameter and even, in some situations, the
initial condition used in the generation of a time-series from a
given chaotic system \cite{piccardi06,wang08}.

\section{Conclusions}
The present work is a summary of different methods based on the
theory of dynamical systems that can be used to point out some
design problems in chaos-based cryptosystems. The main conclusion
derived from the set of tools proposed is that the tests of security
of conventional cryptography are not the only to be considered in
the context of chaotic cryptography. Indeed, either the design or
the analysis of a chaotic cryptosystem must be done along with a
thoroughly knowledge about the dynamics of the chaotic map selected.
Otherwise, the resulting encryption scheme could present serious
security and efficiency problems.

\bibliographystyle{IEEEbib}

\end{document}